\documentclass[aps,prstab,reprint,groupedaddress,nofootinbib,longbibliography]{revtex4-2}

\usepackage{graphicx}
\usepackage{dcolumn}
\usepackage{bm}
\usepackage{amsmath}
\usepackage{units}
\usepackage{lipsum}  
\usepackage[english]{babel}
\usepackage[utf8]{inputenc}
\usepackage{grffile}
\usepackage{booktabs}
\usepackage[normalem]{ulem}

\usepackage[colorinlistoftodos]{todonotes}
\usepackage[caption=false]{subfig}

\usepackage{hyphenat}
\usepackage{placeins}

\usepackage{color}

\newcommand{\newtilde}{\raise.19ex\hbox{$\scriptstyle\mathtt{\sim}$}}

\newcommand{\tmu}{%
  \ifmmode
    \mathchoice
        {\hbox{\normalsize\textmu}}
        {\hbox{\normalsize\textmu}}
        {\hbox{\scriptsize\textmu}}
        {\hbox{\tiny\textmu}}%
  \else
    \textmu
  \fi
}

\begin{document}

\global\long\def\d{\mathrm{d}}
\global\long\def\en{\varepsilon_{0}}
\global\long\def\e{\mathrm{e}}
\global\long\def\ii{\mathbf{\imath}}



\title{Asymmetric Influence of the Amplitude-Dependent Tune Shift on the Transverse Mode-Coupling Instability}



\author{Miriam Brosi}\email{miriam.brosi@maxiv.lu.se}
\author{Francis Cullinan}
\author{Åke Andersson}
\author{Jonas Breunlin}
\author{Pedro Fernandes Tavares}
\affiliation{MAX IV Laboratory, Lund University, Lund, Sweden}


\date{\today}

\begin{abstract}
At the 3\,GeV ring of the MAX IV Laboratory, a fourth generation ring-based synchrotron light source, an asymmetric influence of the sign of the amplitude-dependent tune shift (ADTS) on the transverse mode-coupling instability (TMCI) has been observed. 
Measurements of the instability, in dedicated single-bunch experiments at low chromaticity, revealed a significant dependence of the dynamics of the instability above threshold on the sign of the ADTS. While for a negative sign of the ADTS the crossing of the instability does not lead to a loss of beam current, a positive sign results in the loss of 40\% or more of the beam current at the threshold.
In order to investigate the observed asymmetry, systematic measurement of beam dynamics above the threshold have been conducted in combination with particle tracking simulations with mbtrack2 and theoretical calculations of the Landau damping due to the ADTS. The findings point towards an influence of the Landau damping in combination with the low synchrotron frequency, which indicates that this effect could become relevant in future low-emittance electron storage rings.
\end{abstract}

\pacs{no PACS numbers yet}

\maketitle
\section{Introduction}

The transverse mode-coupling instability (TMCI) in electron storage rings is a single-bunch transverse instability and an important collective effect which can limit the parameter space for stable operation. Especially in fourth generation light sources, this collective effect can strongly influence the achievable operation parameters. The instability depends on many beam parameters like the natural bunch length, the chromaticity and the tunes.
The connection with the amplitude-dependent tune shift (ADTS) was in the past investigated in the interest of mitigation of the instability by Landau damping~\cite{Hereward:1114390, mounet_landau_2020, carver_transverse_2018} at proton storage rings, as the required betatron tune spread can among other sources come from the ADTS.

First studies of the influence of the ADTS on the TMCI for the 3\,GeV ring of the MAX IV Laboratory where presented by several of the authors in~\cite{Tavares:xe5034, LER2018Cullinan}.
While in the case of MAX IV the TMCI does not affect routine operation, it is
nevertheless important to characterize and further investigate such instabilities as with the continuous push towards more extreme operation modes and beam parameters new effects and interaction between parameters can arise and become relevant for the mitigation of such instabilities in future machines.

Dedicated, systematic experiments have now been conducted in single-bunch operation showing an asymmetric influence of the sign of the ADTS on the dynamics of the vertical TMCI above threshold. While the ADTS does not seem to significantly affect the threshold current, it changes the behaviour of the bunch above threshold. For values of the ADTS close to zero, a partial beam loss is observed when the threshold current is crossed while slowly increasing the bunch current.
For ADTS with a large absolute value, on the other hand, beam loss is not observed as the threshold is crossed and the instability leads solely to oscillations of the bunch centre-of-mass and a beam size blow-up.
For a positive sign of the ADTS, this partial beam loss occurs up to higher values of the ADTS than for a negative sign, resulting in a significant asymmetry in the encountered beam loss above threshold.

At ADTS values where no beam loss occurs, the beam dynamics above the TMCI threshold can be studied.
This is used to investigate the observed asymmetry further.
It is observed that, as expected, the instability leads to a blow-up of the bunch size and strong center-of-mass oscillations. For negative ADTS values, and at high current at high positive ADTS values, these oscillations are additionally amplitude-modulated with a much lower frequency showing a sawtooth shaped pattern.

This paper compares measurements conducted at the MAX IV Laboratory with dedicated simulations and theoretical considerations. The measurements include systematic scans of the instability threshold and the occurring current loss as well as time-resolved measurements of the beam dynamics, in this case the center-of-mass motion and the transverse bunch size, above the threshold current.
The simulations consist of particle tracking with the mbtrack2 code~\cite{gamelin:ipac2021-mopab070} which allows the inclusion of the amplitude-dependent tune shift.
Theoretical calculation considering transverse mode-coupling and Landau damping due to the ADTS were conducted as described in the next section and section~\ref{sec:theo_calc}.

\subsection{Transverse Mode-Coupling Instability}\label{sec:theory}

\begin{figure}
\centering
   \includegraphics[trim = 0mm 0mm 0mm 0mm, clip, width=0.5\textwidth]{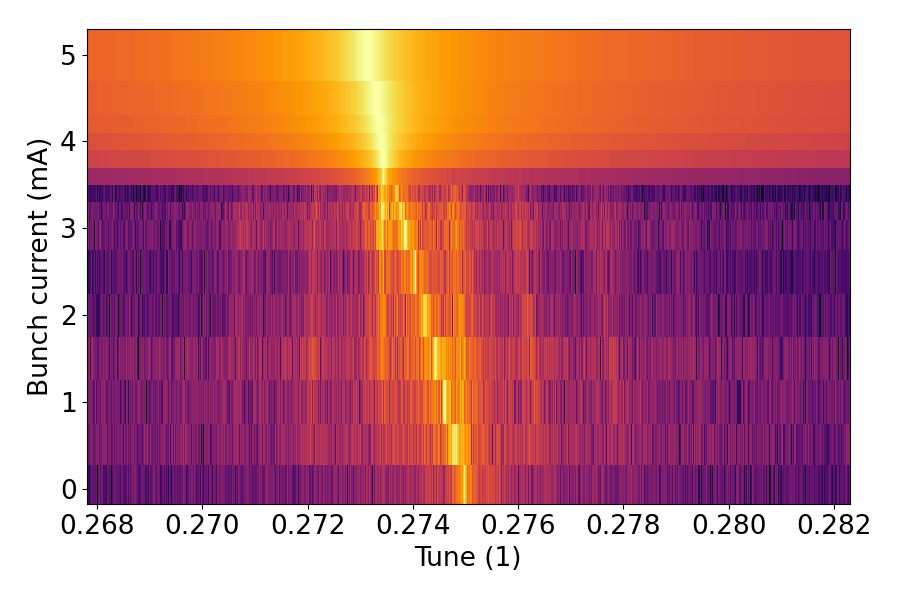}
\caption{Simulated coherent beam spectrum of the bunch at zero chromaticity and zero ADTS, showing the tune shift with increasing bunch current and the resulting mode coupling at the TMCI threshold at 3.5\,mA.}
\label{fig:tuneshift_0adts}
\end{figure}

The transverse mode-coupling instability can arise when the current dependent tune shift due to the transverse impedance leads to a coupling of the coherent betatron tune with one of the neighboring head-tail mode frequencies (typically the mode -1 with a separation of $-\nu_s$, the synchrotron tune).
The TMCI, which occurs at zero chromaticity, has, opposed to the head-tail instability, a well defined threshold current at which the growth rate increases abruptly.
Figure~\ref{fig:tuneshift_0adts} shows the simulated mode coupling at zero chromaticity.

The theory of Landau damping in combination with transverse mode-coupling has been developed by Chin~\cite{chin}. It is characterized by a dispersion integral written as
\begin{equation}\label{eq:dispint}
I_m=-2\pi\int_0^\infty \frac{J}{V-m\nu_s-\nu_\beta-\Delta\nu(J)}\left(\frac{df}{dJ}\right)dJ,
\end{equation}
where $V$ is the complex coherent tune of the instability to be found, $J$ is the action of betatron oscillation, $\nu_\beta$ and $\nu_s$ are the betatron and synchrotron tunes respectively, $f$ is the normalized charge distribution in $J$ of the bunch and $\Delta \nu(J)$ is the amplitude-dependent tune shift (ADTS) giving rise to the tune spread.

For the theoretical calculations a Gaussian charge distribution is assumed:
\begin{equation}
f(J)=\frac{1}{2\pi\langle J\rangle}e^{-\frac{J}{\langle J\rangle}}\, ,
\end{equation}
where $\langle J\rangle$ is the average action of the particle ensemble.
For our purposes, the tune shift is assumed to be proportional to the action $J$ resulting in the following definition where $b$ is the amplitude-dependent
\begin{equation}
    \Delta\nu\left(J\right)=b\cdot J \label{equ:j_equation}
\end{equation}

In the following, when a value for the ADTS coefficient is given, it refers to $b$ and has the unit $\left[b\right]=\nicefrac{1}{\textrm{m}}$, if not stated otherwise.

Evaluating the integral in Eq.~\ref{eq:dispint} gives
\begin{equation}\label{eq:exp1}
I_m=-\frac{1}{b\langle J\rangle}\left[1+\zeta e^{-\zeta}E_1(-\zeta)\right]\;\text{where}\;\zeta>0
\end{equation}
and
\begin{equation}
\zeta=\frac{V-\nu_\beta-m\nu_s}{b\langle J\rangle}\, .
\end{equation}
Solutions with small and positive growth rates $\mathrm{Im}(V)>0$ lie on the boundary of stability. To include radiation damping, we could set $\mathrm{Im}(V)=+1/\tau_x$ where $\tau_x$ is the radiation damping time. Here, however, we neglect radiation damping as it does not make a large difference in the case of the TMCI. An added benefit is that Eq.~\ref{eq:exp1} is then scalable by $b$ and so only needs to be evaluated twice, once for $b>0$ and once for $b<0$.

The inverse of the dispersion relation is subtracted from the diagonal elements of Chin's scaled coupling matrix $\nu_s\mathbf{M}_{nl}^{mk}$ for head-tail and mode-coupling instabilities  as given by Eq.~2.44 in~\cite{chin}. Solutions are determined numerically by equating the determinant of the resulting matrix to zero:
\begin{equation}\label{eq:detzero}
\mathrm{det}(I^{-1}_m\delta_{ml}\delta_{nk}-\nu_s\mathbf{M}_{nl}^{mk})=0
\end{equation}
where $\delta_{ij}$ is the Kronecker delta. In practice, there are two unknowns left to determine: the tune spread at $\langle J\rangle$ ($\Delta\nu(\langle J\rangle)=b\langle J\rangle$) and the coherent frequency of oscillation $\mathrm{Re}(V)$.

\section{Experimental Setup}

\begin{table}[b]
\centering
\caption{Beam Parameters during Measurements}
\label{tab:table_parameters}
\begin{tabular}{l c}
\toprule
 \textbf{Parameter} & \textbf{Value}\\
 \midrule
 Beam energy\,/\,GeV & 3.0\\
 Circumference\,/\,m & 528\\
 RF frequency\,/\,MHz & 99.931\\
 Harmonic number & 176\\
 RF voltage\,/\,kV & 864 \\
 Synchrotron freq.\,/\,Hz & 830 \\
 Synchrotron tune & 0.00146\\
 Vertical tune & 16.275\\
 Vertical beta function\,/\,m &2.0 to 16.1\\
 Horizontal tune & 42.2\\
  Horizontal beta function\,/\,m &0.8 to 9.8\\
 \bottomrule
 \vspace{-7mm}
\end{tabular}
\end{table}

\color{black}
The measurements presented in this paper were performed with the machine parameters given in Table~\ref{tab:table_parameters} if not stated otherwise.
Only a single bunch was stored in the machine to be able to use diagnostics that are not bunch-resolved such as beam position monitors (BPM) and the synchrotron light monitor (SLM). To ensure that only a pure single bunch is filled, great care has been taken to clean out residual charge from the other buckets.
The used BPMs and the used SLM sit at different position in the ring, and therefore at different values of the beta function. To make the bunch sizes and center-of-mass positions shown within this publication comparable, all measurements are scaled to correspond to a measurement at a beta function of 16.0\,m. This value was chosen as it is the beta function at the position of the used SLM and is very close to the maximal vertical beta function of 16.1\,m. To translate the measured values to this position, they are scaled with the ratio of the square roots of the goal beta function value and the original value at the place of measurement.

The total acceleration voltage in the main cavities was set to a fixed value for better comparability of different measurements. Due to the usage of a single bunch and therefore a low absolute beam current, the passive Landau cavities are not elongating the bunch.
Likewise, all insertion device gaps were opened for the sake of comparability.
The reference orbit for the orbit correction was set to zero-orbit without any beamline bumps\footnote{Beamline bumps refer to deliberate deviations in the closed orbit at the position of IDs to help synchrotron radiation alignment for beamlines.} laid in, to use an optic close to the design optics used in the simulations.
During the measurements, the orbit correction was only in use while changing to new settings, e.g. changing chromaticity or ADTS and was afterwards switched off.
When changing settings, it was also checked that the tunes were at the routine working point.
The vertical chromaticity was reduced via sextupole magnets in the vertical plane while the horizontal chromaticity remained at the value for routine operation of around 1.1.
For adjustments of the ADTS three octupole magnet families are available~\cite{Tavares:xe5003_lattice}.

 \begin{figure}
\centering
   \includegraphics[width=0.5\textwidth]{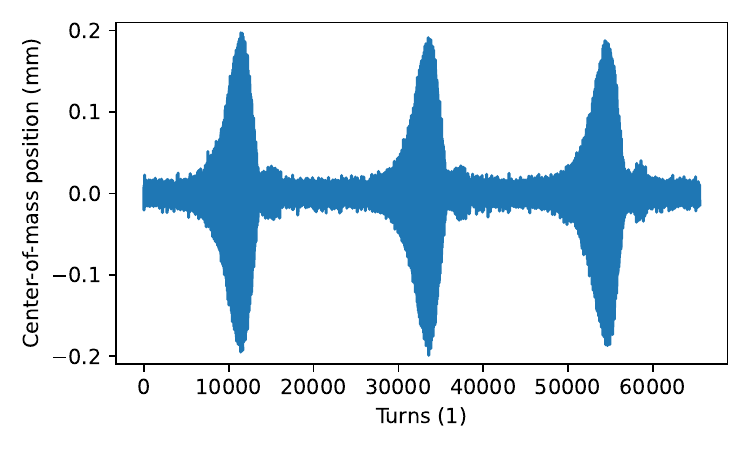}
\caption{BPM signal showing the amplitude modulation caused by the TMCI on the vertical center-of-mass position as function of turns for an ADTS coefficient of $b=-6868/\mathrm{m}$ and 4.5\,mA. 
}
\label{fig:bpm}
\end{figure}

\subsection{ADTS - Measurement and Control \label{sec:adts}}

The center-of-mass (COM) motion is measured via the turn-by-turn data from the BPMs. The position at each turn can be written out for $2^{16}$ consecutive turns. As the measurements where taken in single-bunch operation, this gives the COM position of the bunch at each turn at every BPM.
The coherent tunes can therefore be calculated from the Fourier transform of this data.

\begin{figure}
\centering
   \includegraphics[width=0.5\textwidth]{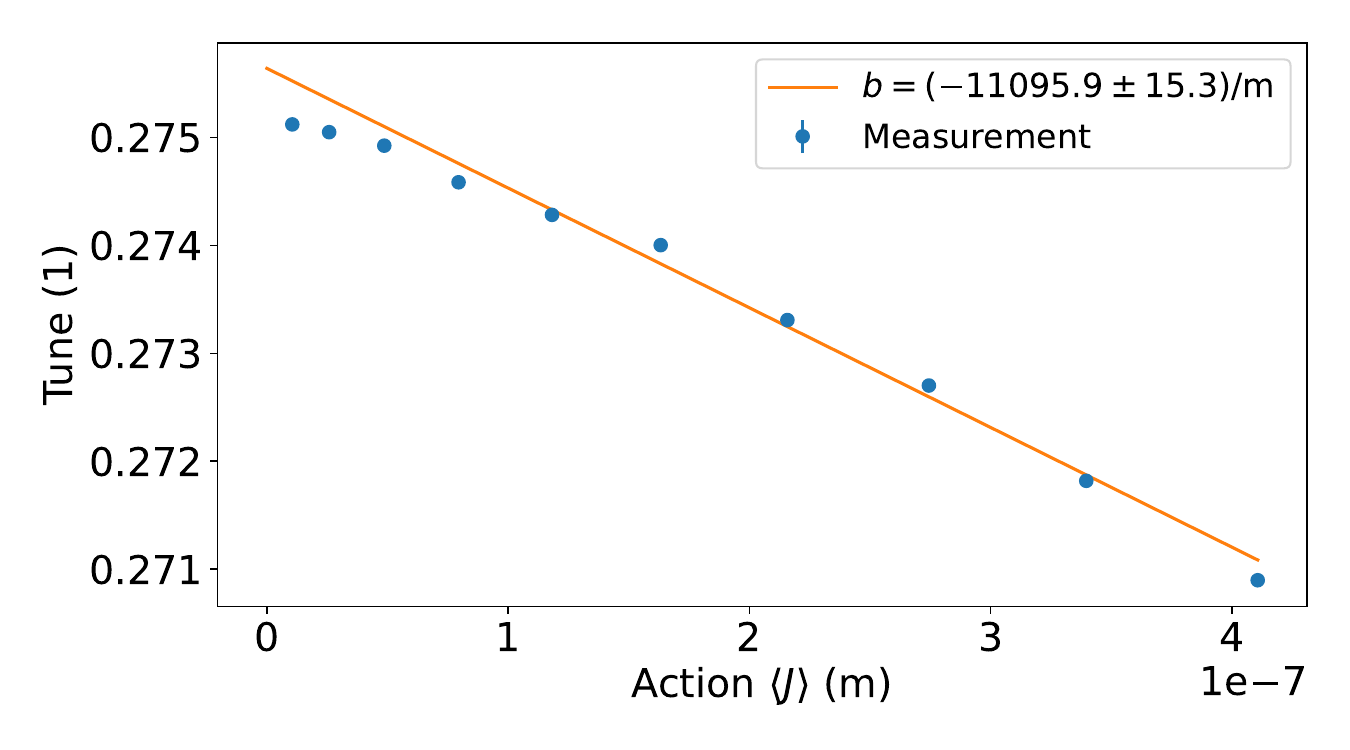}
\caption{Measurement of the betatron tune as function of the action $\langle J\rangle$ in the vertical plane, showing the linear dependence giving the amplitude-dependent tune shift coefficient  $b$.}
\label{fig:adts}
\end{figure}

The amplitude-dependent tune shift was measured by kicking the bunch in each transverse plane individually with increasing amplitude while detecting the center-of-mass movement on the turn-by-turn BPM data for both transverse planes.
The tune was calculated via the Numerical Analysis of Fundamental Frequencies (NAFF) algorithm \cite{Kostoglou:IPAC2017-THPAB044NAFFlib, NAFFlib_github} based on the first 100 turns after each kick.
The resulting tunes for the different kick amplitudes show the tune shift as a function of the center-of-mass displacement $\hat{x}$ at each BPM position.
For the conversion between the measured maximal amplitude $\hat{x}$ at each BPM and the action $J$, it is assumed that at maximum displacement $\hat{x}$ the action $J$ can be calculated via the corresponding value of the beta-function $\beta_\textrm{s}$ at the position of measurement $s$, in this case the position of each BPM:
\begin{equation}
J=\frac{\hat{x}^2_{\mathrm{s}}}{\beta_\mathrm{s}}.\label{equ:conversion}
\end{equation}
Figure~\ref{fig:adts} shows the near-linear dependence of the tune shift on the action $J$ in an example measurement.

To ease the operation during the experimental scans of the ADTS value, a response matrix $M$ was measured for the resulting change of the ADTS caused by changes to two of the octupole families. This matrix allows fast calculation of the necessary change $\Delta I_\mathrm{oct,u}$ in octupole current for a requested change of $\Delta b_\mathrm{u}$ in ADTS coefficient, with $u$ representing the planes $x$ and $y$,  without the need for intermediate measurements.
\begin{equation}
\begin{pmatrix}\Delta I_\mathrm{oct,x}\\ \Delta I_\mathrm{oct,y}\end{pmatrix}
= M^{-1} *\begin{pmatrix} \Delta b_\mathrm{x} \\ \Delta b_\mathrm{y}\end{pmatrix}
\end{equation}
Nevertheless, after arriving at a new ADTS value and checking, and if necessary correcting, the chromaticity and the tunes, and before conducting dedicated measurements, the ADTS value was measured to ensure accuracy.

\subsection{Transverse Bunch Size}
\begin{figure}[!tb]
\centering
   \subfloat[]{
   \includegraphics[trim=5mm 0mm 0mm 0mm, clip, width=0.24\textwidth]{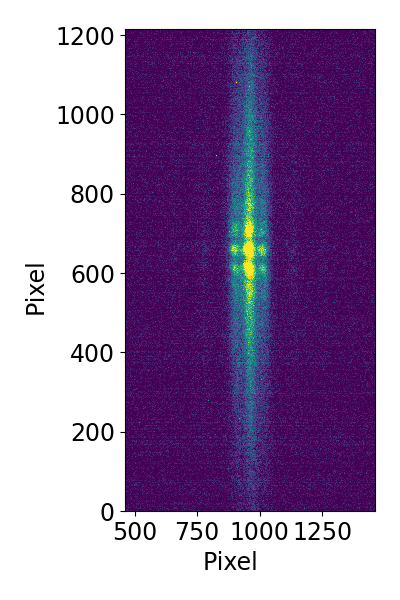}
         \label{fig:foto}}%
   \subfloat[]{
   {\includegraphics[trim=5mm 0mm 0mm 0mm, clip,width=0.24\textwidth]{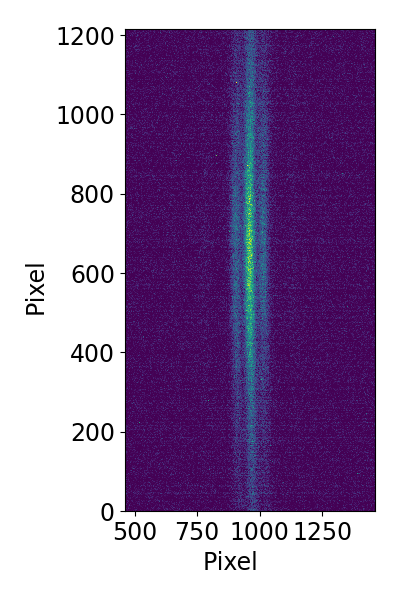}}
   \label{fig:foto2}}
\caption{Synchrotron light spot of a vertically unstable bunch measured at a diagnostic beam line for interferometric bunch size measurement. a) The additional, residual low-current bunches are stable and show the typical interferometric pattern on top of the unstable main bunch. b) After cleaning the residual bunches, only the vertically unstable single bunch is visible.}
\label{fig:fotos}
\end{figure}

The transverse bunch size is measured at the two diagnostic beam lines \cite{Breunlin:IPAC2016-WEPOW034diag} via synchrotron light monitors (SLM) with interferometric source point imaging.
Synchrotron radiation in the visible wavelength range is detected with CMOS cameras after passing though a double-slit for the horizontal plane and a diffraction obstacle in the vertical plane.
The beam sizes can be calculated from the interferometric visibility in the resulting interference pattern~\cite{Breunlin:IPAC2016-WEPOW034diag}. During the instability, the bunch size is blown up to such a degree that the interference pattern is not visible anymore (Fig.~\ref{fig:foto2}) and the distribution is fitted by a Gaussian. 
The required exposure time of the cameras to gather enough intensity, does not allow for turn-by-turn detection.
Nevertheless, the exposure time is short enough ($\approx$ 1\, ms) to allow resolving the time structure of the characteristic amplitude modulation observed on the center-of-mass position (see Fig.~\ref{fig:bpm}) caused by the dynamics above the instability threshold.
In this case, it has to be taken into account that during the exposure time the camera integrates over the observed center-of-mass oscillations providing a superposition of the center-of-mass motion and the interference pattern containing the beam size information.
Furthermore, it has do be asserted that no residual bunches are present during these measurements.
As visible in Fig.~\ref{fig:fotos}, even a small amount of charge in additional residual bunches around the main single bunch, in this case approximately 5\% of the charge, can significantly influence the observed spot profile on the SLMs.
As these low-charge residual bunches are below the instability threshold and therefore stable, their light shows the classical interference pattern. The intensity of the focused pattern is therefore overshadowing the smeared-out spot profile from the unstable main bunch.

\subsection{Synchronised Measurements}

The acquisition by the cameras can be triggered so that synchronised images can be taken relative to the turn-by-turn center-of-mass motion measured with the BPMs.
The synchronization was aligned with triggered kicks from a pulsed magnet to the beam which can be observed in both systems (BPMs and SLMs).
The timing between the camera acquisition and BPM acquisition is chosen such that the camera's exposure time window lies roughly at three-quarters of the BPM measurement window of $2^{16}$ turns ($\approx 115$\,ms). By this, the center-of-mass movement is known for some time before and after the bunch size measurement. The alignment accuracy depends on the camera exposure time used and is in the presented measurements better than $1$\,ms.

In case of multiple such measurement sets being taken during the instability with the characteristic amplitude modulation on the center-of-mass movement (see Fig.~\ref{fig:bpm}), the repetitive behaviour seen on the BPMs can be used to overlay multiple measurement sets aligned by this pattern. This will provide a ``sampled'' image of the changes in the light spot observed on the SLM cameras. In other words, due to the measurement trigger not being synchronised to the instability dynamics, different phases of the amplitude modulation are sampled with every measurement set taken and the repetitiveness of the amplitude modulation can be used to reconstruct a time resolved image.
The spot size is the result of the superposition of the blown up bunch size and the center-of-mass oscillation within the exposure time window.

\section{Simulation tool\label{sec:simu}}

To simulate the beam dynamics observed, especially above the threshold current, we performed particle tracking with the mbtrack2~\cite{gamelin:ipac2021-mopab070} python code.

A broadband resonator was used for the vertical impedance with a shunt impedance of 200\,kOhm/m at the resonant frequency of 11.5\,GHz and a quality $Q$ of 1~\cite{SKRIPKA2016221transImp}.
The mbtrack2 simulations also included a longitudinal impedance (732\,Ohm at 6\,GHz with $Q=1$~\cite{Skripka:NAPAC2016-WEA3CO04_longImp}) to account for bunch lengthening with increasing bunch current.
mbtrack2 allows for the optics parameter to be read-in from an AT lattice file using pyAT. The RF voltage was set to the same value as in the measurements (see Table~\ref{tab:table_parameters}) and synchrotron radiation effects were included in the simulations.
The tune shift contribution by the ADTS is calculated in mbtrack2 based on the action $J$ (see~Sec.~\ref{sec:theory} Eq.~\ref{equ:j_equation}).
Intra-Beam Scattering (IBS) is not yet implemented in mbtrack2.
The simulations were conducted with 50000 macro-particles and were run on the COSMOS cluster of LUNARC at Lund University.
The center-of-mass amplitudes and bunch sizes are calculated in mbtrack2 once per turn and for a value of the beta function of 7.23\,m corresponding approximately to the average of the beta function along the ring. Therefore, all simulated COM amplitudes and bunch sizes were scaled to the same reference as the measurements, a beta function of 16.0\,m.

 \begin{figure}
\centering
     \includegraphics[width=0.48\textwidth]{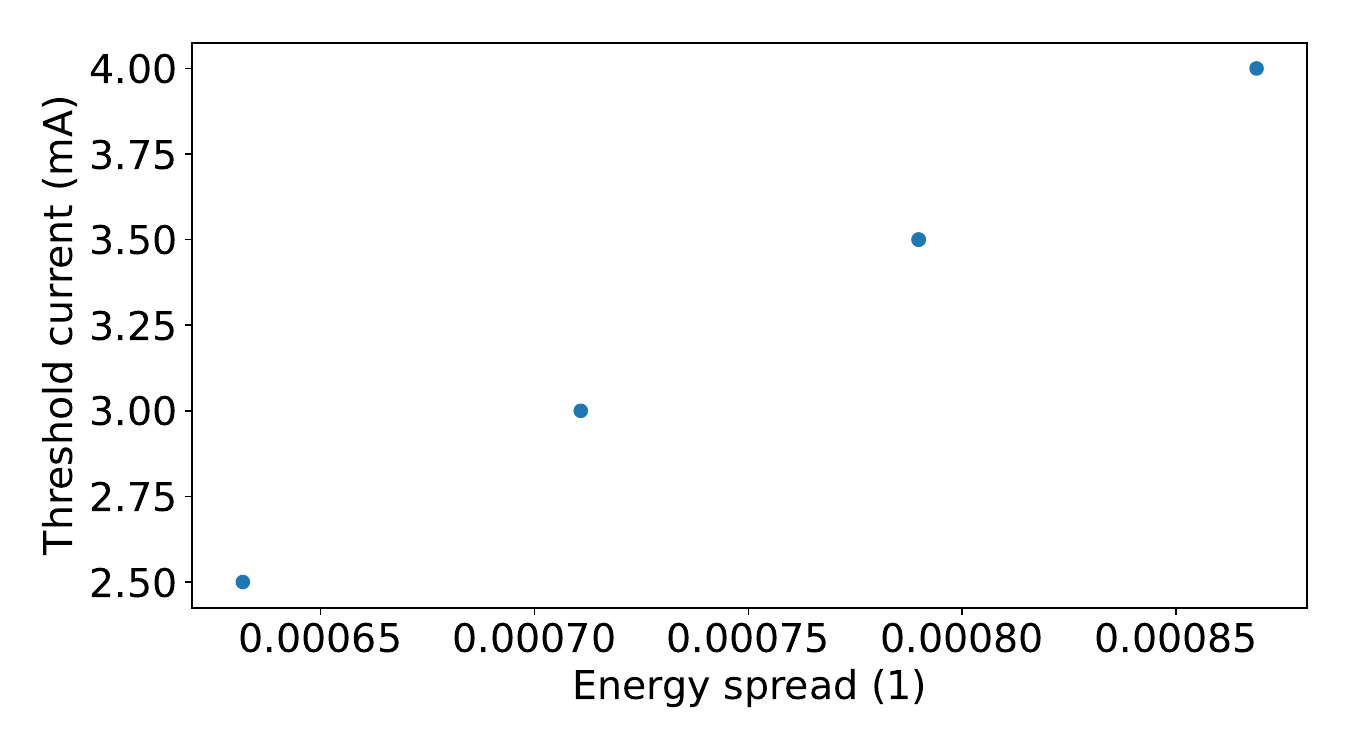}
\caption{Simulated TMCI threshold at a chromaticity of $0.05$, $b=1000$ and different energy spreads of 80\%, 90\%, 100\% and 110\% showing the expected increase in threshold for increased energy spread.}
\label{fig:espread_scan}
\end{figure}

\begin{figure*}
\centering
    \includegraphics[width=0.5\textwidth]{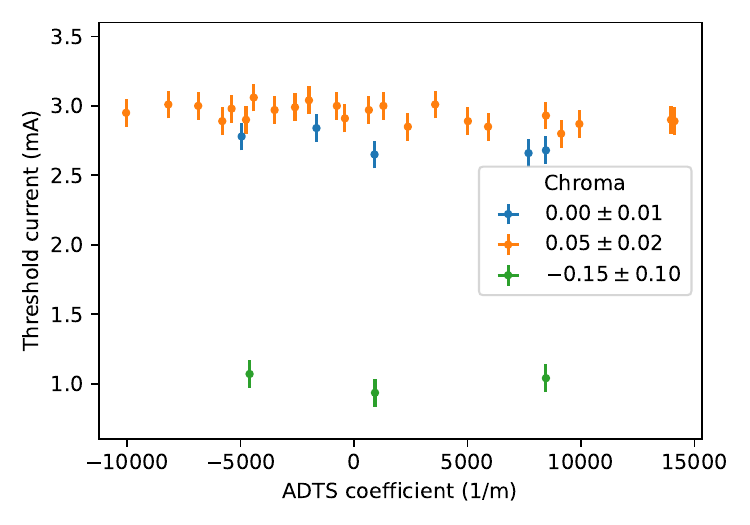}\includegraphics[trim=0mm 0mm 0mm 0mm, clip, width=0.5\textwidth]{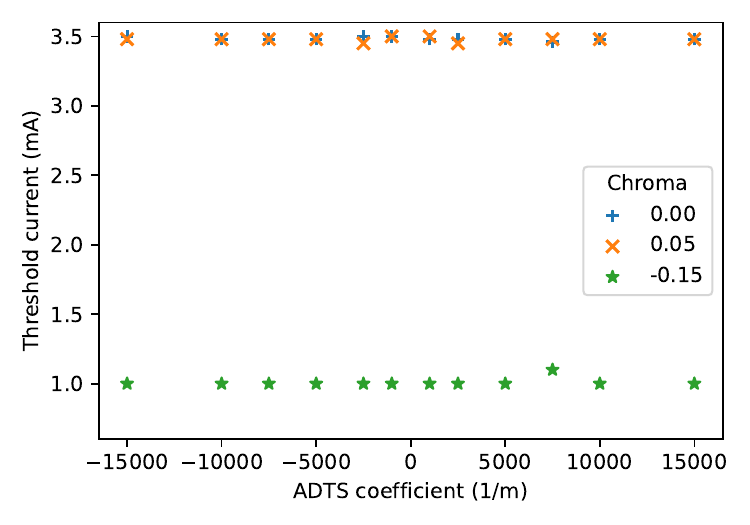}
\caption{left: Single bunch threshold currents during injection shown as a function of ADTS coefficient measured at chromaticities of 0, 0.05 and -0.15.
right: Single bunch threshold currents simulated in mbtrack2 (including bunch lengthening by addition longitudinal impedance) for chromaticities of 0, 0.05, -0.15.}
\label{fig:injthr}
\end{figure*}

\section{results}
In the following the measurement and simulation results are presented side by side and grouped by the different beam properties affected by the instability. The measurements were conducted in the vertical plane.
Besides the threshold current, the beam loss at threshold, the bunch position and size as well as the betatron tune shift with current below and above the instability threshold are discussed. Additionally, theoretical calculations on the Landau damping in combination with the transverse mode-coupling instability will be discussed in the context of the observed asymmetry with respect to the sign of the amplitude-dependent tune shift.

\subsection{Instability Threshold\label{sec:threshold}}

When studying an instability, the threshold current is a very important parameter as it is the limit up to what current stable operation is possible.

We observed that the TMCI threshold current changes depending on the beam conditions while reaching the threshold. For example, the threshold current was lower when the beam current slowly decayed while the beam was unstable, compared to the threshold current when charge was injected into a stable beam. Additionally, within a certain bunch current range, it was possible to stabilize an unstable beam with the bunch-by-bunch feedback system and 
the beam remained stable after switching off the feedback. Furthermore, the instability could be triggered by excitations or kicks to the beam even below the injection threshold, but not below the decaying threshold.
In summary, we observed an hysteresis effect on the TMCI threshold current, where a stable beam shows a higher threshold current than an already unstable or excited beam.

A possible explanation for the threshold hysteresis is Intra-Beam Scattering.
For beams with small transverse emittances, IBS can lead, amongst other things, to an increase in energy spread. This can be mitigated by increasing the vertical emittance either via coupling or with vertical excitations of the beam.
With respect to the TMCI, IBS would have the following effect. For a stable beam the vertical emittance is small and the energy spread is increased by IBS. An increased energy spread results in an increase of the theoretical TMCI threshold, as the current-dependent tune shift is inversely proportional to the bunch length which again is proportional to the energy spread \cite{Chao:1993zn}\color{black}.
As soon as the beam becomes unstable, either by crossing the (higher) threshold or by excitation, the vertical emittance increases and the effect of the IBS is reduced leading to a reduction in energy spread. The lower energy spread finally results in a lower TMCI threshold current. This results in a hysteresis of the instability threshold depending on whether the threshold is measured starting with a stable or an unstable beam.

As the mbtrack2 simulations do not include IBS this hysteresis can not be directly simulated. Nevertheless, simulations with the energy spread manually set to different values show the expected dependence of the threshold current on the energy spread (see Fig.~\ref{fig:espread_scan}).

For the following studies, the threshold during injection was selected as it can be quickly measured reliably and accurately compared to the other thresholds. Furthermore, the disturbance to the stored beam caused by the Multipole Injection Kicker (MIK) is known to be very small~\cite{ALEXANDRE2021164739MIK}, so the observed threshold during injection should be very close to the theoretical threshold (including IBS) if the charge in a stable beam is slowly increased.

The left hand side of Fig.~\ref{fig:injthr} shows the thresholds measured during injection for different ADTS coefficients. These thresholds where determined by injecting (using the MIK) into a single bunch and observing the center-of-mass movement on the BPMs. As soon as the center-of-mass movement grew unstable the injection was stopped and all charge in residual bunches from a non-perfect single bunch injection was cleaned.
The resulting threshold currents differ as expected depending on the chromaticity. To separate the TMCI from the head-tail instability \cite{ng_physics_2006}, the measurements where conducted either at a vertical chromaticity of zero or nearly zero chromaticity (0.05) in contrast to a chromaticity of $\approx 1.1$ during routine operation. The thresholds for both chromaticity values are very similar and lie around 2.8\,mA. Additionally, measurements were conducted at a slightly negative vertical chromaticity of $-0.15$. As expected during operation with a positive momentum compaction factor and a negative chromaticity (e.g.~\cite{Chao:1993zn}), they show a much lower threshold current of around 1\,mA.
The same is visible in the simulated thresholds shown on the right hand side of Fig.~\ref{fig:injthr}. The simulated thresholds for a chromaticity of zero and 0.05 lie both at around 3.45\,mA and are higher than seen in the measurements by about 0.5\,mA. At the same time, the simulated threshold for the slightly negative chromaticity matches the measurements at around 1\,mA.

The measurements and the simulations were conducted for a range of positive and negative ADTS coefficients. No significant correlation between threshold currents and the value of the ADTS coefficient is observed in either measurement or simulations.
This is not unexpected as the experimental ADTS coefficients reached only result in a very small tune shift for the center-of-mass oscillation and the bunch size observed in a stable beam. A typical measured ADTS coefficient of $b=5000/\mathrm{m}$ leads with a stable bunch size or a center-of-mass movement with a maximal amplitude of less than $10\,\mathrm{\tmu m}$ (corresponding to an action of approximately $2\cdot10^{-11} \mathrm{m}$) to a tune shift in the order of only $\Delta\nu\approx10^{-7}$. Consequently, an ADTS coefficient in this order of magnitude is not relevant until the instability starts to blow-up the beam leading to a bigger contribution of the ADTS due to the then drastically increased center-of-mass oscillation and bunch size.
As is shown in Fig.~\ref{fig:bpm}, during the instability the center-of-mass amplitudes reach values of the order of hundreds of micrometers and, as will be shown later, the bunch size blows up to similar sizes. Then the tune shift by ADTS is in the order of $\Delta\nu\approx0.001$ which corresponds already to two-thirds of the synchrotron tune.
Therefore, it is then, above the instability threshold, that the ADTS is expected to influence the dynamics.

\subsection{Beam Losses at Threshold}
\begin{figure}
\centering
   \includegraphics[width=0.5\textwidth]{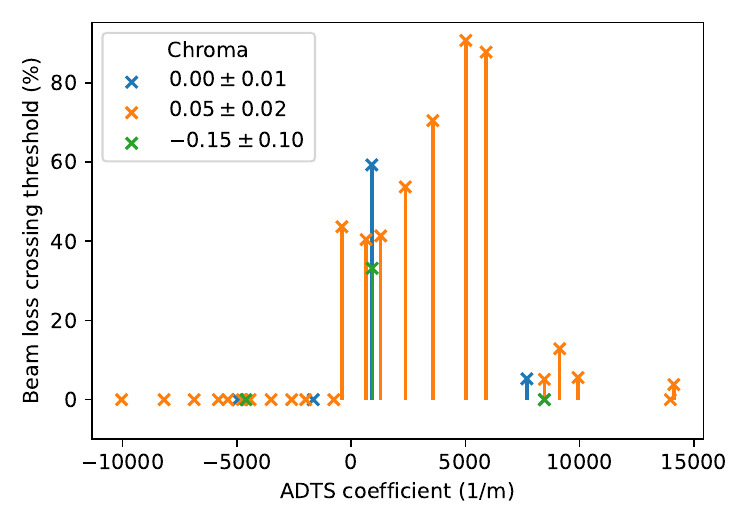}
\caption{Current loss in percent at the TMCI threshold as function of the amplitude-dependent tune shift  at chromaticities of 0, 0.05 and -0.15.} \label{fig:thrloss}
\end{figure}

A significant influence of the magnitude and sign of the ADTS coefficients can be observed in the amount of charge lost when the instability threshold is crossed during injection. Figure~\ref{fig:thrloss} shows the beam loss in percent for different values of the ADTS coefficient with a chromaticity close to zero or with slightly negative chromaticity values. For negative ADTS coefficients up to nearly zero ($\approx -500 /\mathrm{m}$) no beam loss is encountered at all when crossing the instability threshold during injection.
This is already noteworthy as it shows that the instability is not destructive even though it leads to strong center-of-mass oscillations and an increase in bunch size.
On the other side, for positive ADTS coefficients a partial beam loss is observed when crossing the threshold. For values from zero up to $6000 /\mathrm{m}$ more than 40 and up to 90 percent of the beam current is lost. For higher positive ADTS coefficients, the loss goes down close to zero again.
So, for ADTS coefficients up to  $6000 /\mathrm{m}$, there is a difference in the observed behaviour for a positive and negative sign of the ADTS coefficient. While at negative coefficients the instability is self-containing, for positive coefficients a partial beam loss is observed until the beam stabilizes again.

To investigate this difference in behaviour above the threshold the time domain signal of the center-of-mass oscillation and the bunch size was studied in measurement and simulation.

\subsection{Bunch Position and Size \label{sec:timedomain}}
 \begin{figure}
\centering
     \includegraphics[width=0.5\textwidth]{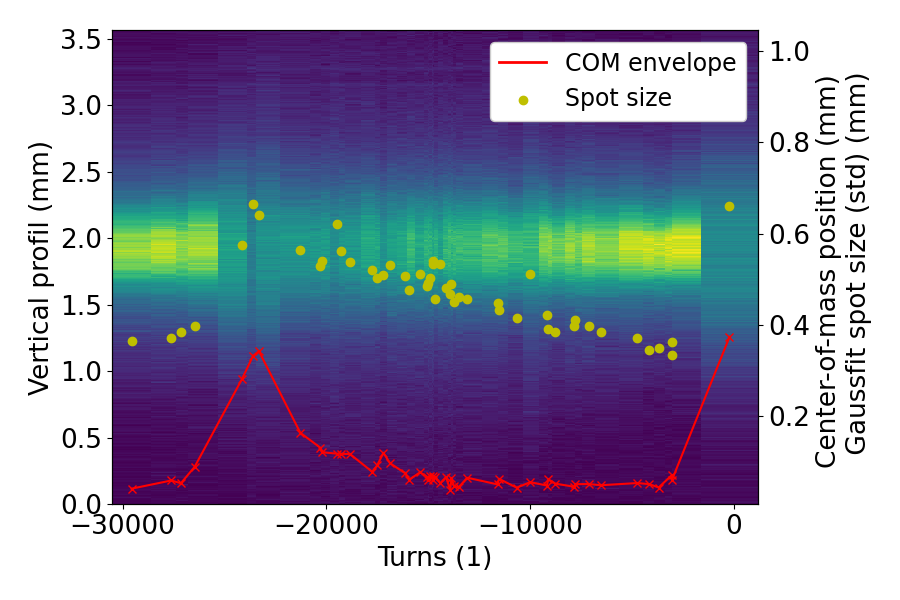}
\caption{Synchronized measurement of vertical spot profile and  center-of-mass amplitude as function of time. The image shows the vertical light  spot measured at different points in the sawtooth like dynamic. In red the envelope of the center-of-mass motion is displayed with a point for each spot profile measurement. The yellow dots indicate the spot size (std) gained from a Gaussian fit. Measured at ADTS coefficient $b=-10000/\mathrm{m}$ and current of 2.6\,mA. \label{fig:sync_single}}
\end{figure}

\begin{figure}
   \includegraphics[width=0.5\textwidth]{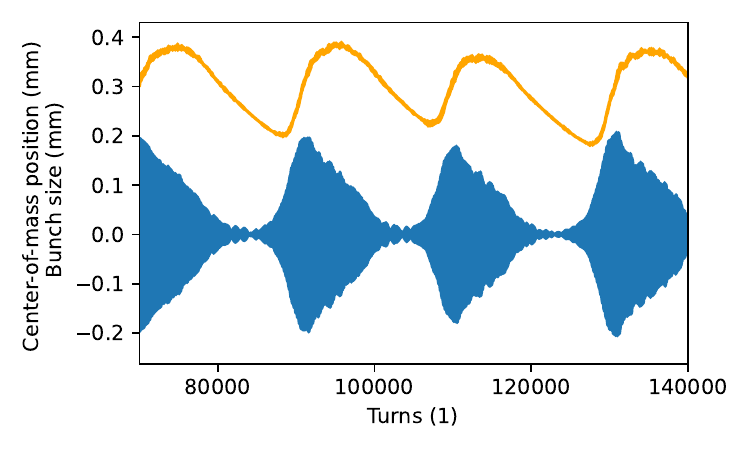}

\caption{Simulated center-of-mass oscillation and bunch size at an ADTS coefficient $b=-15000/\mathrm{m}$ and a current of 4.8\,mA.}
\label{fig:com_sim}
\end{figure}

For a negative ADTS coefficient, the dynamic above threshold shows clear, regular, sawtooth like bursts in transverse bunch size and as amplitude modulation of the center-of-mass oscillations. This is visible in the BPM trace directly (Fig.~\ref{fig:bpm}) as well as in the synchronous measurement of bunch position and bunch size in Fig.~\ref{fig:sync_single}.
The contribution of the bunch size can be seen in the difference in behavior over turns between the vertical beam-size and centre of mass. Additionally, the calculated spot size of up to 0.68\,mm (standard deviation determined by Gaussian fit) in Fig.~\ref{fig:sync_single} compared to the detected center-of-mass oscillation amplitude of maximal 0.4\,mm (= 0.8\,mm peak-peak) indicates that the bunch size has a non-negligible contribution.
Furthermore, the center-of-mass oscillation goes back to nearly zero for some hundred turns between the increases in oscillation amplitude. During this time the observed spot size slowly damps down indicating that the bunch size is damping down, reaching minimal values in the order of 0.35\,mm before the next peak.
The same sawtooth behaviour is present in the simulation shown in Fig.~\ref{fig:com_sim} when synchrotron radiation effects, such  synchrotron radiation losses, synchrotron radiation damping and quantum excitations are included.

\begin{figure*}
\centering
   \includegraphics[trim=0mm 0mm 0mm 0mm, clip,width=0.5\textwidth]{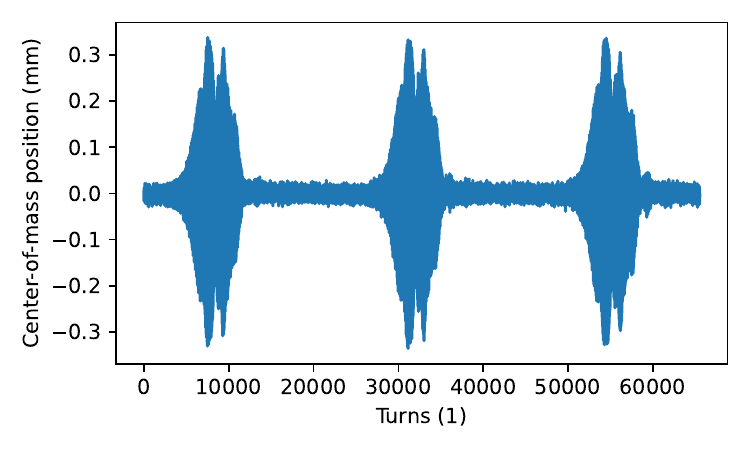}%
   \includegraphics[trim=0mm 0mm 0mm 0mm, clip,width=0.5\textwidth]{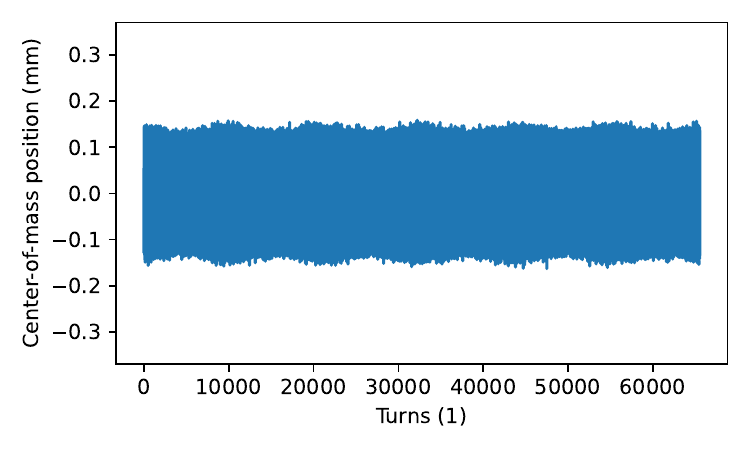}

      \includegraphics[trim=0mm 0mm 0mm 0mm, clip, width=0.5\textwidth]{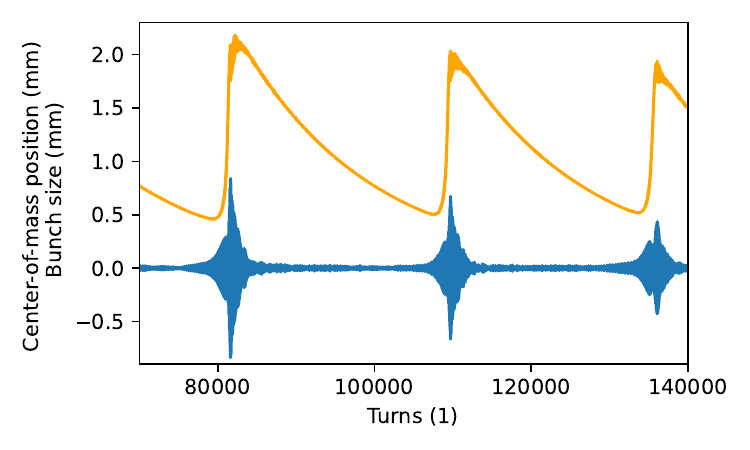}%
      \includegraphics[trim=0mm 0mm 0mm 0mm, clip,width=0.5\textwidth]{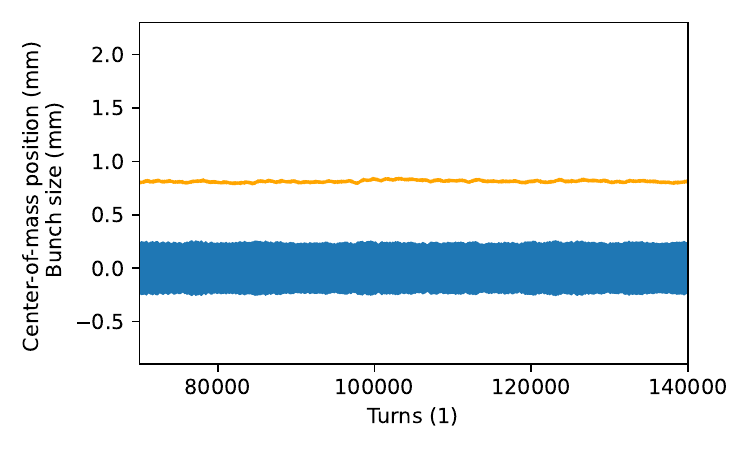}

\caption{Top: Measured center-of-mass oscillation at 3.07\,mA (left) and 2.79\,mA (right) and an ADTS coefficient $b=13720$/m.  The measurement at high current show a similar sawtooth pattern to measurements at negative ADTS coefficient (Fig.~\ref{fig:bpm}). The measurement at low current is more constant (noise level would be +-0.02\,mm). Bottom: Simulated center-of-mass oscillation and bunch size at $b=15000$/m and bunch currents of 4.8\,mA (left) and 4.6\,mA (right).}
\label{fig:com_size_posadts}
\end{figure*}

This dynamic indicates a stabilizing mechanism which leads to a containment of the instability instead of a continuous growth until charge is lost. The behavior in the amplitude of the center-of-mass oscillations and the bunch size show that at one point a temporary stabilization occurs which leads to a damping of the oscillation to below the noise limit of the measurement. The bunch size is also damped down during this stable period but it does not reach the expected stable bunch size before the instability is triggered again leading to a fast blow-up of the bunch size and the onset of strong center-of-mass oscillations.
A possible mechanism behind these dynamics, is Landau damping. The blown-up bunch size, increased COM oscillations and the ADTS lead to an increased tune spread which can cause a temporally increased Landau damping effect.
The result can be described as a ``self-containing'' instability, caused by Landau damping, which only sets in when the bunch is blown-up and the ADTS results in a significant tune shift. Two points can be identified in the dynamics. An ``upper turning point'' when the damping due to the ADTS caused by the blow-up of the beam becomes predominant and overpowers the growth driven by the instability. And a ``lower turning point'' when the bunch size and COM oscillations are damped down to such low values that the resulting ADTS is not enough to Landau damp the instability any longer and the growth starts again.

While the IBS is considered as the cause of the hysteresis observed in the Instability threshold, it is not a candidate for explaining the self-containing dynamics, as it operates in the wrong sense, ie. a blown-up beam has a lower threshold current and is therefore more unstable and does not contribute to a self-containing effect, where the threshold would need to increase to temporarily stabilize the unstable beam.

For positive ADTS, the dynamics above the threshold can only be observed in measurements at high ADTS coefficients where no instantaneous charge loss occurs. For higher bunch currents, the dynamics in the center-of-mass oscillation and the bunch size have a similar sawtooth like pattern as observed for negative ADTS coefficients (see left side of Fig.~\ref{fig:com_size_posadts}). The pattern changes for lower bunch currents, closer to the threshold, as shown on the right side in Fig.~\ref{fig:com_size_posadts}. Here, the center-of-mass oscillation amplitude is only lightly modulated in the measurements, and nearly constant in the simulations.

\begin{figure}
\centering
   \includegraphics[width=0.5\textwidth]{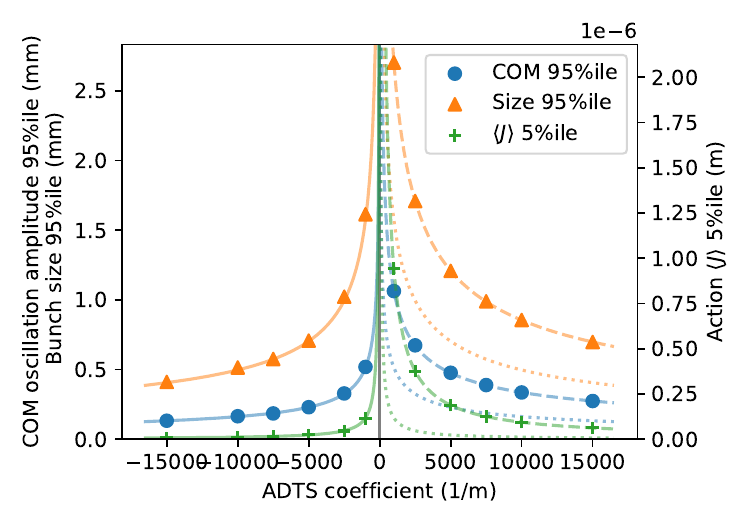}
\caption{Simulated maximal (95th percentile) bunch size and center-of-mass oscillation amplitude and minimal (5th percentile) action $\langle J\rangle$ as a function of the ADTS coefficient for currents slightly above threshold. The lines highlight the $1/x^{1/2}$ (bunch size and COM) respective $1/x$ ($\langle J\rangle$) dependency with the dotted line being the mirror of the solid line at negative ADTS.
}
\label{fig:saturation_sim}
\end{figure}

\begin{figure}
\centering
   \includegraphics[
   width=0.5\textwidth]{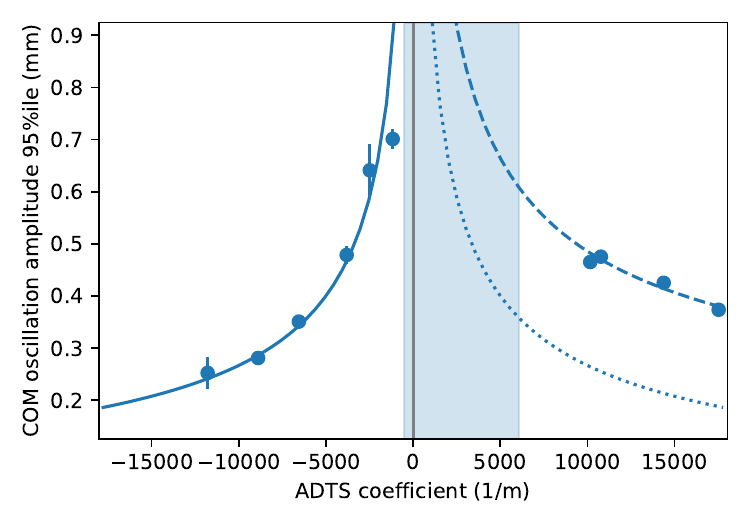}
\caption{Measurements of the maximal center-of-mass oscillation amplitude (95th percentile) as a function of the ADTS coefficient. The measurements were taken at bunch currents close to the threshold and with a chromaticity of 0.05. The lines highlight the $1/x^{1/2}$ dependency with the dotted line being the mirror of the solid line at negative ADTS. The errors show the standard deviation between multiple consecutive measurements per point.}
\label{fig:ampl_over_adts_current2}
\end{figure}

The simulation results show that a positive  ADTS (Fig.~\ref{fig:com_size_posadts}, lower left plot) leads to a stronger blow-up  than a negative ADTS (Fig.~\ref{fig:com_sim}). The experiments (Fig.~\ref{fig:bpm} and Fig.~\ref{fig:com_size_posadts}, upper left plot) show the same  asymmetric behaviour
This indicates, that the level of blow-up at which the instability is contained (the upper turning point) and finds some kind of equilibrium, pseudo-stable state is different for negative and positive ADTS.
The asymmetry is clearly visible in Fig.~\ref{fig:saturation_sim} where, for a bunch current slightly above threshold, the maximal bunch size and the maximal oscillation amplitude of the center-of-mass is given as a function of ADTS coefficient in simulations\footnote{To be more robust against outliers the 95th percentile of the bunch size and the center-of-mass oscillation amplitude are taken.}.
The range in ADTS coefficient where partial current loss occurs depends on the combination of the center-of-mass oscillations and the total bunch size which, above a certain value, leads to parts of the charge being ``scraped'' by the beam pipe.
Independent of the exact value, it can be seen from Fig.~\ref{fig:saturation_sim} that the affected ADTS range would not be symmetric around zero but rather shifted to positive ADTS coefficients.
The measurements show the same dependence of the maximal center-of-mass oscillation amplitude\footnote{Again, the 95th percentile of center-of-mass oscillation is taken.} as a function of the ADTS coefficient (Fig.~\ref{fig:ampl_over_adts_current2}) for negative ADTS. While it is not measurable at the lower positive ADTS coefficients, due to the partial beam losses, the measured values at higher positive ADTS coefficients are higher than the corresponding values at negative ADTS showing the same asymmetry as in the simulations in Fig.~\ref{fig:saturation_sim}. So to reach the same level of suppression of the instability, meaning low values in maximal bunch size and center-of-mass oscillations, a higher positive than negative ADTS coefficient is be needed. This asymmetry observed in both measurement and simulations provides the experimental explanation for the asymmetry observed in the beam loss at the threshold (see Fig.~\ref{fig:thrloss}).

\begin{figure}
\centering
   \includegraphics[trim=0mm 3mm 0mm 0mm, clip,width=0.5\textwidth]{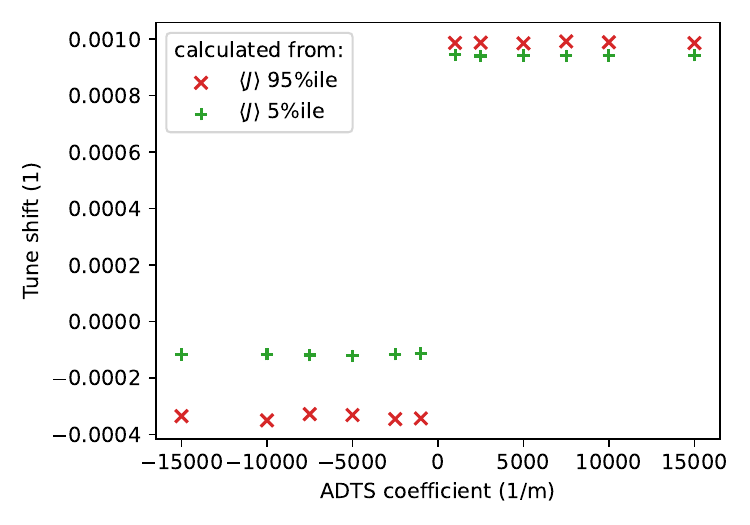}
\caption{Tune shift calculated from the maximal (95th percentile) and the minimal (5th percentile) action $\langle J \rangle$ as function of ADTS coefficient.}
\label{fig:tune_shift_caused_by_maxJ}
\end{figure}
Figure~\ref{fig:saturation_sim} also shows the value of the average action of the particle ensemble $\langle J \rangle$ at the times of minimal bunch size and center-of-mass oscillations\footnote{The 5th percentile is taken as value for the minimal action $\langle J \rangle$.} as a function of ADTS coefficient, again the asymmetry for the different signs of the ADTS coefficient is visible. The minimal value $\langle J \rangle$ reaches can be connected to the point were the instability is no longer damped and the beam becomes unstable again, the lower turning point of the dynamics.

\begin{figure*}
\captionsetup[subfloat]{captionskip=4pt}
	 \subfloat[Negative ADTS]{
\includegraphics[width=0.48\linewidth]{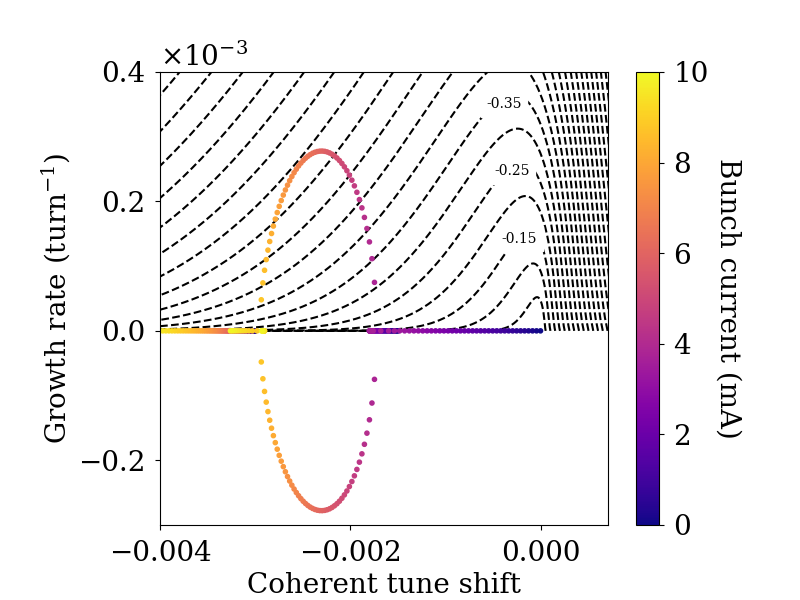}}
\subfloat[Positive ADTS]{
\includegraphics[width=0.48\linewidth]{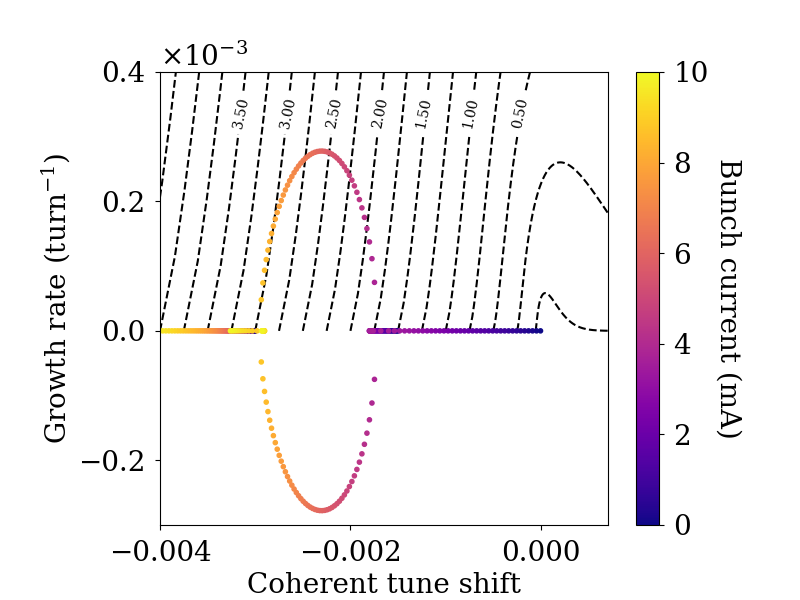}}
%
\caption{Stability diagrams for negative ADTS coefficient (left) and positive ADTS coefficient (right). The numbers that label the Landau contours indicate the magnitude of the tune shift at $\langle J\rangle\times10^{3}$. The colored points are the eigenvalues of the scaled coupling matrix $\nu_s\mathbf{M}_{nl}^{mk}$ where the color represents the bunch current.}\label{fig:stabdiags}
\end{figure*}

Both, the maximal bunch size and center-of-mass oscillation amplitudes as well as the minimal action $\langle J \rangle$ show the characteristic dependence on the ADTS coefficient. For  $\langle J \rangle$ it follows a $1/x$ dependency and the bunch size as well as the center-of-mass oscillation amplitude has a $1/\sqrt{x}$ dependency. This again indicates the connection with the tune shift, via Eq.~\ref{equ:j_equation}. Going one step further and calculating the tune shift due to ADTS from the simulated values of the minimal action $\langle J \rangle$ for each ADTS coefficient, shows a constant but different level of tune shift for each sign of the ADTS (Fig.~\ref{fig:tune_shift_caused_by_maxJ}). While for negative ADTS the calculated tune shift of $\approx0.00012$ is approximately 8\% of the synchrotron tune, the shift for positive ADTS is with $\approx0.00094$ already 65\% of the synchrotron tune.
This significant difference implies, that for positive and negative ADTS a different level of tune shift is required to contain the instability via sufficient Landau damping. The tune shift calculated from the maximal  $\langle J \rangle$ shows a very similar behavior, where the small difference between maximal and minimal values on the positive side of the ADTS is explained by the fact that nearly no sawtooth is observed for positive ADTS coefficients and currents close to the threshold (compare Fig.~\ref{fig:com_size_posadts}).
We conclude that the tune shift stays between these two levels, with the higher one being the point were the beam stabilises and starts to damp (upper turning point) and the lower one indicates when the stabilizing effect stops and the beam goes unstable and blows up again (lower turning point). And while the overall behaviour is the same for both signs of the ADTS, the tune shift at the turning points is significantly higher for positive ADTS coefficients.

That this can be attributed to Landau damping, is supported in the following section by theoretical calculations comparing the instability growth rates of the transverse mode coupling against Landau contours.

\subsection{Theoretical Calculations}\label{sec:theo_calc}

Figure~\ref{fig:stabdiags} shows stability diagrams calculated from Eq.~\ref{eq:exp1} for the cases of positive and negative ADTS. The calculations were performed with an internal code which was bench marked against DELPHI~\cite{delphi}.
Contours are drawn by plotting the imaginary part of the inverse dispersion relation $I^{-1}_m$ against the real part. These contours map out an asymmetric shape in complex frequency space pointing towards negative coherent tune shifts in the case of negative ADTS. Changing the sign of the ADTS to positive reflects the contour about the line of zero coherent frequency shift. Also shown on the figure are the eigenvalues of the coupling matrix in Eq.~\ref{eq:detzero} without Landau damping where three modes are included: the azimuthal head tail modes $m=0$, $m=-1$ and $m=-2$. As radial modes have been neglected in order to make the image clearer, it can be observed (contrary to measurements) that the beam becomes stable again at higher currents. For a head-tail mode to be stable in isolation, its complex coherent frequency shift would have to be within the Landau contour.
The condition in the presence of mode coupling is slightly different and given by the zero determinant in Eq.~\ref{eq:detzero}.
How these diagrams are best applied to the TMCI to solve  Eq.~\ref{eq:detzero} is an active research topic, discussed for example in~\cite{mounet_landau_2020}.
Nonetheless, the images in Fig.~\ref{fig:stabdiags} are still illustrative. It can be seen that, in order to influence the stability, a positive ADTS coefficient must be much larger in magnitude than a negative one, matching the observations in measurement and simulation shown in Fig.~\ref{fig:saturation_sim}~and~\ref{fig:ampl_over_adts_current2}.
This is intuitive as a negative ADTS coefficient means that the tune spread is towards negative tune shifts and the current-dependent tune shift of the $m=0$ mode is also negative for most broadband impedances. Therefore, the shifted coherent tune frequency lies within the band of the incoherent tune and motions with this tune are damped by the Landau damping effect.

\begin{figure*}
    \captionsetup[subfloat]{captionskip=1pt}
	 \subfloat[Simulation at negative ADTS]{
	\includegraphics[width=0.5\textwidth]{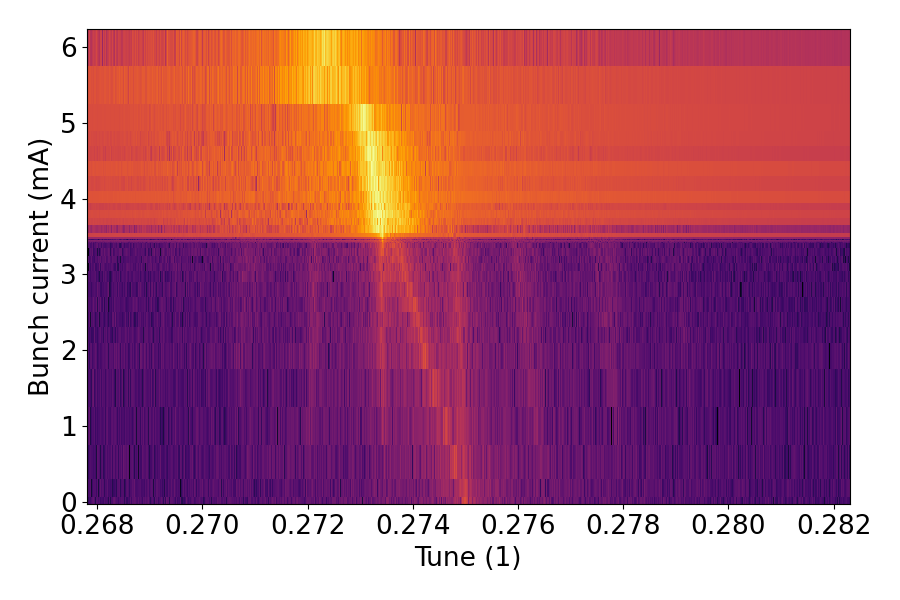}\label{fig:simu_modes_neg}}
	\subfloat[Simulation at positive ADTS]{
	\includegraphics[width=0.5\textwidth]{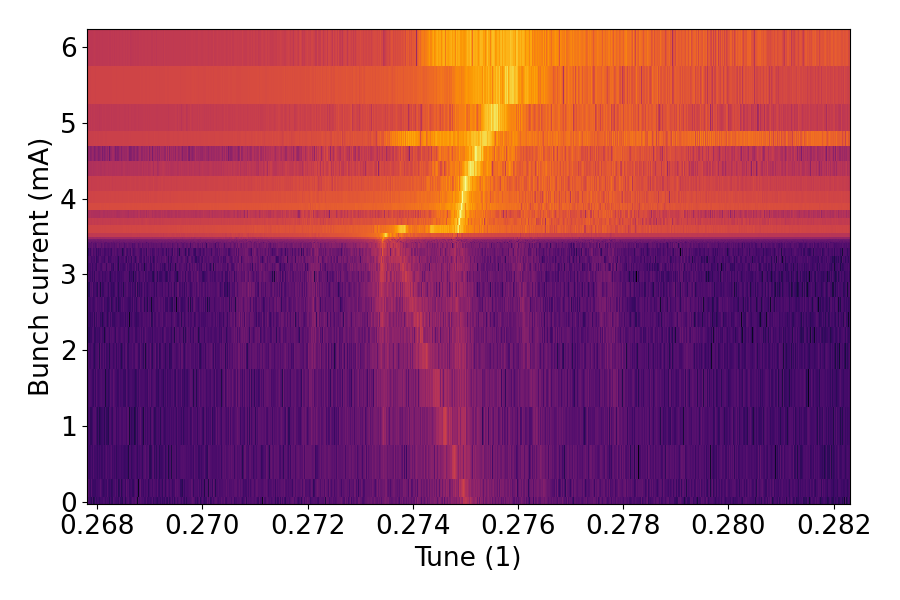}\label{fig:simu_modes_pos}}\\

    \subfloat[Measurement at negative ADTS]{
	\includegraphics[width=0.5\textwidth]{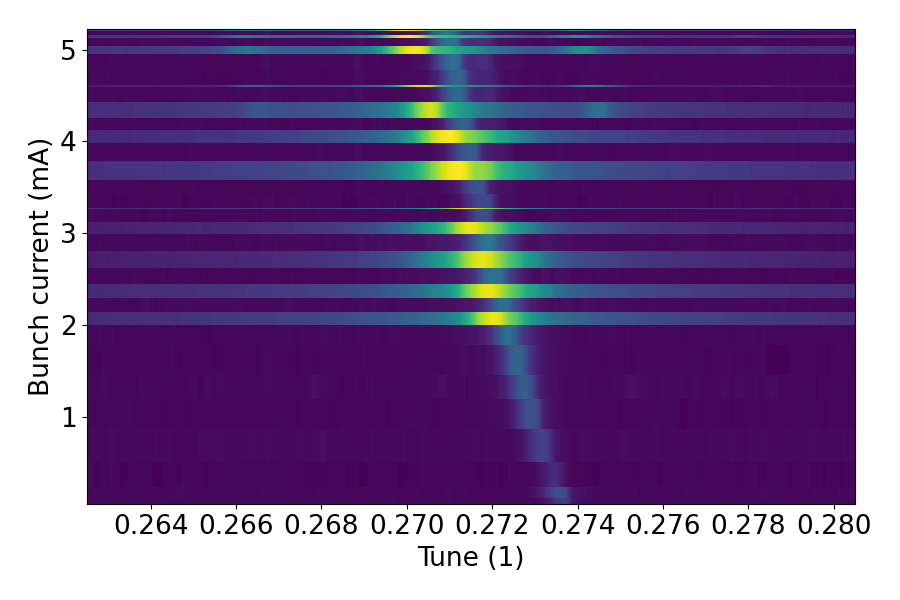}\label{fig:meas_modes_neg}}
	\subfloat[Measurement at positive ADTS]{
	\includegraphics[width=0.5\textwidth]{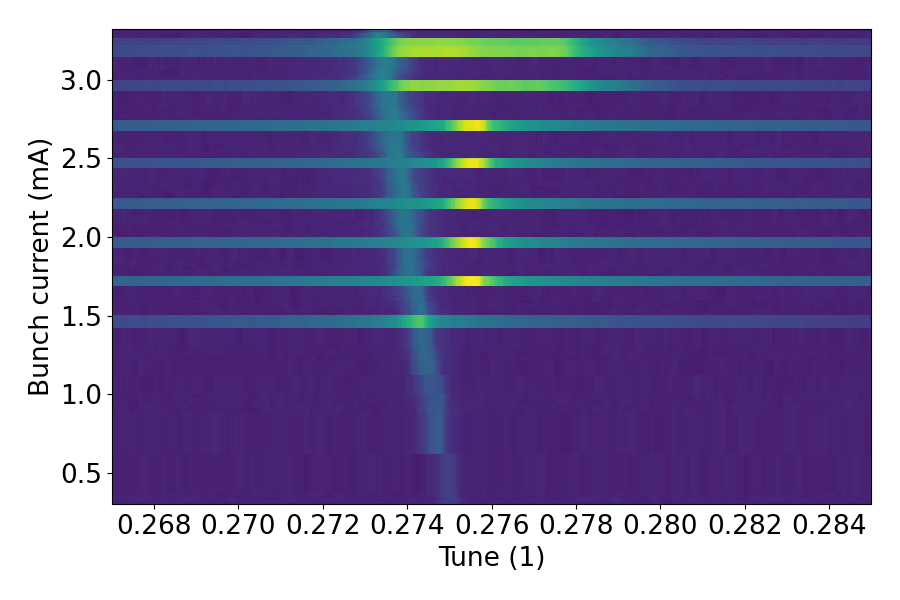}\label{fig:meas_modes_pos}}	\\

	\vspace{-0.15cm}
\caption{Coherent motion spectrum showing the current-dependent betatron tune shift below and above the instability threshold. Simulation (top): Fourier transform of the center-of-mass oscillation plotted as a function of the bunch current for negative (left) and positive (right) ADTS coefficient $b=15000$/m.
Measurement(bottom): Fourier transform of the center-of-mass oscillation
as a function of bunch current for an ADTS coefficient of $b=-10000$/m (left) and $b=13720$/m. During the measurement the instability was ``switched'' on and off.}
	\label{fig:modes}
\end{figure*}

One feature of storage rings used for fourth-generation synchrotron light sources, particularly those using low-RF frequencies (such as the 100~MHz of the 3~GeV ring at MAX~IV), that is beneficial in this regard is the low incoherent synchrotron frequency. This means that the coupling frequency of the $m=0$ and $m=-1$ head-tail modes is not so far out of the spread in betatron tune of the electron bunches. Nevertheless, both in simulation and measurement at the 3~GeV ring at MAX~IV, the magnitude and sign of the ADTS coefficient does not impact the threshold current of the TMCI. As discussed in Sec.~\ref{sec:threshold}, this is because the point of the mode-coupling is outside the tune spread of the bunch when it is stable. When the bunch goes unstable and increases in size, however, the tune spread increases until the Landau damping kicks in.

\subsection{Betatron Tune Shift with Current}

Observing the vertical betatron tune as function of current directly shows the expected current-dependent tune shift due to the transverse impedance from the zero-current tune of $\nu_{0}=0.275$ towards the -1 mode at the first synchrotron frequency side band ($\nu_{0}-0.00146$). In simulations of the coherent beam spectrum the threshold of the TMCI is clearly visible as the current at which the tune couples to the -1 mode (top row in Fig.~\ref{fig:modes}). This is the same for both signs of the ADTS.
The difference for negative and positive ADTS starts above the threshold current where for negative ADTS the tune continues its shift towards a lower tune with a similar slope as below the threshold (Fig.~\ref{fig:simu_modes_neg}). For the positive ADTS, the behaviour looks very different. While a slight shift in the opposite direction to higher tunes would not be unexpected, due to the positive sign of the ADTS, the tune jumps within a very small current range above the threshold from the -1 mode back to the 0 mode and then shows a continuous shift to higher tunes from there (Fig.~\ref{fig:simu_modes_pos}).

This drastic difference in behaviour can also be seen in measurements. Figures~\ref{fig:meas_modes_neg}~and~\ref{fig:meas_modes_pos} show the measured tune spectra at different bunch currents and for negative and positive ADTS, respectively.
The measurements were conducted in such a way that the previously described hysteresis of the instability threshold (Sec.~\ref{sec:threshold}) was used to get comparative measurements for the tune of a stable and an unstable beam.
To this end, the measurement was started at high bunch currents and the tune spectrum was recorded alternately for a beam stabilized by the BBB feedback system\footnote{After initial stabilization the feedback is switched off during the measurement.} and for an unstable beam, where the instability was triggered by a short excitation\footnote{The excitation is switched off as well before the measurement is taken.}.

For the stable beam, the tune continues its current-dependent shift towards lower values.
For both signs of the ADTS, it is clearly visible that the presence of the instability shifts the tune compared to the tune of the stable beam.
For negative ADTS (Fig.~\ref{fig:meas_modes_neg}), the shift is small and towards slightly lower tune values.
For positive ADTS (Fig.~\ref{fig:meas_modes_pos}), the tune is shifted back towards the zero-current tune (0 mode) and shows a very small current-dependent shift towards higher tune. Except for the difference in threshold and the threshold hysteresis observed in the measurements, the simulation and the measurements agree qualitatively very well with respect to the tune shifts below and above threshold.

At higher bunch currents, additional features appear. In the measurement at negative ADTS, an upper and lower sideband shows up, moving with the tune as function of current.
In the case of the positive ADTS the tune peak is broadened greatly and nearly spans from the -1 mode to the +1 mode.
Comparing with the calculated tune shift of $\approx0.001$ resulting from the maximal action $J$ simulated in case of positive ADTS (Fig.~\ref{fig:tune_shift_caused_by_maxJ}), shows that the jump of the coherent betatron tune by one synchrotron tune ($\nu_{s}=0.00146$) towards the 0 mode is only slightly bigger.
Overall, from the measurement and tracking simulations it is not apparent whether this difference in the behaviour of the coherent tune above threshold is the cause or a consequence of the observed asymmetry in the level of beam blow-up for negative versus positive ADTS.

\section{summary and conclusion}

Landau damping has been investigated in the past as a possible mitigation mechanism of mode-coupling instabilities, also in connection with the amplitude-dependent tune shift as the source of the required tune spread, mainly at proton machines.
In this paper, we showed how Landau damping can affect the dynamics of the transverse mode-coupling instability under certain operational parameters at a fourth generation synchrotron light source.
While during routine operations the bunch current at the 3\,GeV ring at the MAX IV Laboratory is below the TMCI threshold, an asymmetric dependence on the sign of the ADTS has been previously observed in dedicated experiments.
Systematic studies were now conducted to investigate this observed asymmetry in dedicated single bunch experiments.
It was observed, that for some ADTS coefficients the beam was lost when crossing the threshold while at others a saw-tooth shaped amplitude modulation was observed on the center-of-mass oscillation as well as on the bunch size leading to a self-contained instability.

The presented simulations with the tracking tool mbtrack2 and the conducted measurements are in good agreement.
Both show that the observed threshold current is independent of the ADTS coefficient and an observed hysteresis in the measured threshold can be attributed to intrabeam scattering effects.
For the dynamics above the threshold, both measurements and simulation, show that for positive ADTS coefficients the maximal center-of-mass oscillation amplitude and bunch size, that is reached before the instability stabilizes, is systematically higher than for negative ADTS coefficients, indicating that this could be the cause of the observed partial beam current losses.
The same asymmetry is also visible in the tune shift at the minimal $\langle J \rangle$ required for damping. The shift is constant and the value is only dependent on the sign of the ADTS, with the tune shift calculated for positive ADTS coefficients already being at $\approx65\%$ of the synchrotron tune.
Stability diagrams with Landau contours, calculated to include the amplitude-dependent tune shift as well show that higher positive ADTS coefficients compared to negative ones are required for the instability to be Landau damped.
Furthermore, simulations and measurements of the coherent tunes as a function of the bunch current, show a strong difference in the tunes development above threshold. For negative ADTS coefficients the tune is slightly shifting to lower values starting from the -1 mode at the threshold. In contrast, for positive ADTS coefficients, the coherent tune jumps back to the 0 mode and only then shows with increasing current a slight shift to higher tune values, as might be expected for positive ADTS coefficients.

Our experiments and analysis show an asymmetry of how strongly the vertical TMCI is self-containing and of which tune shift with current is observed above the threshold.
As expected, a higher absolute value of the ADTS coefficient leads to a lower maximal center-of-mass oscillation and bunch size blow-up. But when comparing signs, a higher positive ADTS coefficient is required for the same amount of suppression of the instability than for a negative ADTS.

Compared to previous studies on using Landau damping together with ADTS to mitigate the TMCI, two differences were found for the presented investigations for the parameters at the 3\,GeV ring at MAX IV.
It is clear that, the Landau damping only comes in as a stabilization mechanism after the beam has become unstable, as the threshold was shown to not be dependent on the ADTS coefficient, resulting in the observed saw-tooth pattern.

The second result and rather interesting finding is that the maximal level that the center-of-mass motion and bunch size reaches before being contained by Landau damping is higher for a positive than a negative sign of the ADTS coefficient. This could not only be shown in the measurement but also in tracking simulations and in theoretical stability considerations including Landau damping and the ADTS as well. The intuitive explanation behind this is that the ADTS contributes asymmetrically to the Landau damping depending on the ADTS sign,  as the resulting tune spread either covers the same lower frequencies as the tune shift with current by impedance or it goes in the other direction towards higher frequencies reducing the resulting Landau damping effect significantly as seen in the presented results for positive ADTS coefficients.

A key contribution for this asymmetry to become relevant and in making it observable, might be the rather low synchrotron frequency at the 3\,GeV ring. The low momentum compaction factor, typical for fourth-generation light-source storage rings, is combined with the low RF frequency of 100\,MHz. Both of these aspects lead to a synchrotron frequency that then lies within the betatron tune spread of the bunch when it is blown up by the instability. This therefore puts the betatron tune at the point of mode-coupling within the reach of the Landau damping before the beam hits the vacuum chamber wall and is lost.
The role of the low synchrotron frequency in enabling these dynamics indicates that the presented findings will become more relevant in future fourth-generation light-source storage rings with even more extreme parameters than MAX IV.

\begin{acknowledgments}

The computations were enabled by resources provided by LUNARC.

\end{acknowledgments}

\appendix*

\bibliography{bib_tex_TMCI_prab}

\end{document}